\documentclass[%
reprint,
superscriptaddress,
amsmath,amssymb,
aps,
prstper,
]{revtex4-2}

\usepackage{graphicx}
\usepackage{dcolumn}
\usepackage{bm}
\usepackage[T1]{fontenc}	
\usepackage[latin9]{inputenc}	
\usepackage{geometry}		
\usepackage{graphicx}
\usepackage[above,below]{placeins}	
\usepackage{times}
\usepackage[colorinlistoftodos]{todonotes}
\usepackage{hyperref}  
\usepackage{enumerate}
\usepackage{enumitem}
\usepackage{multirow}
\usepackage{booktabs}
\usepackage{xparse}   
\NewDocumentCommand{\rot}{O{45} O{1em} m}{\makebox[#2][l]{\rotatebox{#1}{#3}}}%
\newcolumntype{x}[1]{>{\centering\arraybackslash\hspace{0pt}}p{#1}}

\usepackage{lineno}

\begin{document}
\preprint{APS/123-QED}

\title{Designing Experiments: Student Learning Experience and Behaviour in Undergraduate Physics Laboratories}
\author{Bei Cai}
\email{bei.cai@carleton.ca. Current address: Department of Physics, \\ Carleton University, Ottawa, Ontario, K1S 5B6, Canada}
\affiliation{Department of Physics, Engineering Physics, and Astronomy, Queen's University, Kingston, Ontario, K7L 3N6, Canada}
\author{Lindsay A. Mainhood}
\email{lindsay.mainhood@queensu.ca}
\affiliation{%
Faculty of Education, Queen's University, Kingston, Ontario, K7M 5R7, Canada}
\author{Ryan Groome}%
\affiliation{Department of Physics, Engineering Physics, and Astronomy, Queen's University, Kingston, Ontario, K7L 3N6, Canada}
\author{Corinne Laverty} 
\affiliation{Center for Teaching and Learning, Queen's University, Kingston, Ontario, K7L 3N6, Canada}
\author{Alastair McLean}
\email{mcleana@queensu.ca}
\affiliation{Department of Physics, Engineering Physics, and Astronomy, Queen's University, Kingston, Ontario, K7L 3N6, Canada}

\date{\today}
\begin{abstract}
We investigated physics students' learning experience and behaviour in a second-year laboratory by analyzing transcribed audio recordings of laboratory sessions. One student group was given both a problem and procedure and asked to analyze and explain their results. Another was provided with only the problem and asked to design and execute the experiment, interpret the data, and draw conclusions. These two approaches involved different levels of student inquiry and they have been described as guided and open inquiry respectively. The latter gave students more opportunities to practice ``designing experiments,'' one of the six major learning outcomes in the recommendations for the undergraduate physics laboratory curriculum by the American Association of Physics Teachers (AAPT). Qualitative analysis was performed of the audio transcripts to identify emergent themes and it was augmented by quantitative analysis for a richer understanding of students' experiences. An important finding is that significant improvements can be made to undergraduate laboratories impacting both student learning experience and behaviour by increasing the level of inquiry in laboratory experiments. This is most easily achieved by requiring students to design their own experimental procedures.
\end{abstract}

\maketitle

\section{Introduction}
As physics is an experimental science, it is natural that physicists receive some of their education in a laboratory environment where personal observation and physical experimentation can take place. AAPT identifies the foremost goals of physics laboratories as learning to think like a physicist, referred to as ``habits of mind,'' and constructing a view of the physical world through experimental design, data collection, analysis, and interpretation~\cite{AAPT2015}. The AAPT recommendations for the undergraduate physics laboratory curriculum outlines six learning outcomes that underpin achievement of these goals; these are: constructing knowledge, modeling, designing experiments, developing technical and practical laboratory skills, analyzing and visualizing data, and communicating physics.  

The traditional mode of instruction in physics laboratories frequently uses the ``recipe'' approach where students are given a detailed procedure. This has recently been described as a form of passive teaching, masquerading as active engagement~\cite{Fraser2014}. When students follow detailed instructions without having to grapple with their own conceptual understanding, they are not challenged to think for themselves. Fraser \emph{et al.} argue that, despite the fact they are using physics apparatus, they are solely ``hands-on'' and not ``heads-on''~\cite{Fraser2014}.   

There is emerging interest in student-driven inquiry laboratories that allow students to take ownership of their experimental work~\cite{Aparicio2019, Dounas2017}. The inquiry approach provides opportunities to build conceptual understanding and scientific skills through active participation in design and decision-making processes. Inquiry-based learning is endorsed as learner-focused pedagogy that reinforces achievement of learning outcomes including subject content and thinking skills. It has been shown to improve knowledge of subject content~\cite{Luckie2004} and science process skills~\cite{Arantika2019}. More specifically, it aids conceptual understanding, analysis of experimental errors, interpretation and representation of data, and evaluation of results as well as enjoyment of learning~\cite{Husnaini2019, Holmes2015}.  

A review of 138 inquiry studies in elementary-secondary education concluded that inquiry is a means to prompt active thinking, increase conceptual understanding, and draw conclusions from data~\cite{Minner2010}. A meta-analysis of 22 studies over a ten-year span on the same population confirmed that inquiry prompts active engagement in higher-order thinking skills, such as proposing and evaluating results, which helps students understand science~\cite{Furtak2012}. In a comparison of different levels of inquiry, Spronken-Smith and Walker~\cite{Spronken2010} found that inquiry questions serve as a trigger for learning, student collaboration, teacher facilitation, and increased self-directed learning~\cite{Spronken2010}.

Although the inquiry approach is championed in many post-secondary science programs, characterization of the attributes of inquiry has proven difficult. This may in part be due to the fact that the term is used to describe both a teaching and learning approach as well as a process of investigation~\cite{Carver2012, Buck2008}. 

An inquiry scale, derived from an analysis of nearly 400 undergraduate experiments in 22 lab manuals, was recently introduced by Buck and colleagues~\cite{Buck2008}. They posited that inquiry is a continuum. Within this continuum they defined five levels, each with a different amount of student independence. As the amount of independence increases, the amount of information provided by the instructor decreases. Our focus is on guided and open inquiry. Guided inquiry provides the research question and procedures, and the students are expected to analyze and explain their results. Open inquiry provides the research question and it is up to the students to design and execute the experiment, interpret the data, and draw conclusions.

The purpose of this paper is to provide an evidence-based description of the student learning experiences in two types of inquiry experiments that are part of a second-year undergraduate physics laboratory.  In one experiment a detailed procedure was given, while the other experiment provided the same apparatus and experimental questions, but asked students to develop their own procedure through written prompts. We followed the inquiry scale defined by Buck~\cite{Buck2008} and, although not an exact match, the closest levels of inquiry are \emph{guided} and \emph{open} inquiry, respectively. The guided-inquiry experiment relied heavily on a lab manual for procedures and analysis. The open-inquiry experiment focused on experimental design. While it included the problem with the onus on students to establish procedures and generate explanations of their data, prompts in the lab manual were used to scaffold the learning experiences. Both experiments focused on process rather than product and engaged students in the stages of design, experimentation, and analysis, with varying levels of support.

We begin by describing the course context for the guided- and open-inquiry experiments. The methodology section outlines the research approach and methods for qualitative data collection and analysis. The results discuss the major themes that capture students' learning processes and affective learning behaviours during guided- and open-inquiry experimental experiences. The discussion section compares the different student experiences during the two experiment types. Finally, we offer considerations for the design of physics experiments based on our findings.
 
\section{Course Context}
The second-year undergraduate physics laboratory at Queen's University comprises eleven experiments whose subject matter overlaps the syllabus of the second-year lecture courses: electromagnetism, waves and vibrations, modern physics, and dynamics. 

The first three weeks of the twelve-week semester are devoted to  tutorial laboratories that each cover a specific AAPT learning outcome~\cite{AAPT2015} shown in parenthesis: keeping a laboratory record in a Jupyter Notebook~\cite{JupyterNotebook} (practical laboratory skills), calculating experimental uncertainties using Python and the QExPy Python package~\cite{MARTIN2019100273} (analyzing data), making measurements with an oscilloscope (technical laboratory skills) and the creation of reports using \LaTeX\ and Overleaf (communicating physics). In the remaining nine weeks the students have to complete six experiments and they do that in groups of two. In each three-week period, the groups perform two experiments and, in one of the weeks, they analyze their results or repeat measurements.

The lab manuals inherited from previous instructors were well-written and contained step-by-step procedures in recipe-like~\cite{Silverman1995,AAPT2015} format. Using the scheme introduced above, we would classify them as guided-inquiry experiments~\cite{Buck2008}.

As part of a TRESTLE (multi-institution consortium funded by the National Science Foundation) intervention~\cite{trestlewebsite} and laboratory-redesign project that took place between 2016 and 2019, two of the experiments were redesigned and a new experiment on coupled oscillators was designed and implemented. All three experiments required the students to practice open inquiry. Although the experimental task was still defined by the instructor, the students were required to design their own procedures. 

The experiments selected for redesign were Young's modulus and compact disk (CD) diffraction. In the Young's modulus experiment, the students measured the Young's modulus of steel using both a static and a dynamic method. The CD diffraction experiment, which is the focus of this paper, was introduced to the second-year laboratory course by AM, one of the authors of this paper, circa 1993. The original version of this experiment was based on a short paper by Kettler~\cite{Kettler1991}. Students were given: a helium-neon laser, a CD, and a copy of Kettler's paper containing the grating equation. They were asked to estimate the grating spacing (also called line spacing) of the CD as precisely as possible using their own procedures. Consequently, in its original form, this was an open-inquiry experiment. In subsequent years, other instructors had extended the scope of the experiment by asking the students to use their estimate of the grating spacing to estimate the wavelength of a second laser, and the lab manual had become more guided.

Re-conceptualizing guided-inquiry experiments as open-inquiry experiments can, in some cases, involve relatively minor modifications to the lab manual. Frequently, detailed procedural instructions are replaced with a request for the students to design their own strategy or to evaluate a number of different strategies and choose one. The apparatus, of course, is configured to support at least one experimental strategy and the configuration conveys information about how the instructor expects the experiment to be done. Consequently, we found it necessary to add degrees of freedom (DOF) to support open inquiry. For example, in the CD diffraction experiment it is now possible for students to: change the distance between the CD and the wall where the diffraction pattern is measured, rotate the CD, rotate the laser, and also use a laser with a different wavelength. To avoid overwhelming the students with choices, the lab manual describes the options that are available and the instructor and teaching assistant can help the students make informed choices.  

From the instructor's perspective, the following differences between the guided- and open-inquiry student experiences have been noted: 
(1) Students performing open-inquiry experiments devote time at the beginning of the laboratory period to the design of the experiment and, consequently, they start taking measurements later than those performing guided-inquiry experiments. However, both groups finish the experiment during the three-hour laboratory period. (2) In students' reports who perform open-inquiry experiments, after reflecting on different procedural approaches, they do suggest how the apparatus might be improved. In some cases we have been able to implement these suggestions. (3) When students are given apparatus with multiple DOF they occasionally design experiments which combine DOF in ways that were not necessarily anticipated by the instructor. (4) We found that when students are given the choice of selecting an experiment for a final or culminating report (data from the 2019 session), the open-inquiry experiments are selected, on average, more frequently than the other experiments.

\newcolumntype{L}[1]{>{\raggedright\let\newline\\\arraybackslash\hspace{0pt}}m{#1}}
\begin{table*}[t]
   \caption{Comparison of guided- and open-inquiry manuals and learning outcomes for the CD Diffraction experiment. Students are provided with recipe-like procedures in the guided-inquiry manual while they are prompted to design some of the experimental activities themselves in the open-inquiry manual.} \vspace{0.2cm}
   \label{tab:outlineComp} \centering
   \begin{tabular}{ L{4.2cm} L{3.2cm} | L{5cm} L{4.3cm}  }  \hline 
   \multicolumn{2}{c|}{Guided-Inquiry Experiment} & \multicolumn{2}{c}{Open-Inquiry Experiment}\\
    Manual & Learning Outcomes & Manual & Learning Outcomes \\\hline
    \footnotesize{Make sure that the CD is parallel to the wall and the green laser is perpendicular to the CD; the laser light will be reflected off the CD back into the laser.} & \footnotesize{Students will align the experimental setup by following instructions.} & \footnotesize{Measure all the physical quantities you need to predict what the diffraction pattern should look like with the CD in the vertical geometry and the laser light incident horizontally.} & \footnotesize{Students will decide what physical properties to measure, make the measurements, and use the results to predict where the diffracted light will intercept the wall.} \\ \hline

    \footnotesize{Measure the distance from the CD to the wall. Measure the 1st and 2nd order diffraction spots.} & \footnotesize{Students will make accurate measurements of physical quantities.} & \footnotesize{Establish procedures for ensuring that the CD is parallel to the wall and the laser light is horizontal. Measure the diffraction pattern.} & \footnotesize{Students will design experimental procedures to align the apparatus, decide what quantities to measure, and make the measurements.} \\ \hline

    \footnotesize{Calculate line spacing $d$.} & \footnotesize{Students will calculate $d$ using experimental data.} & \footnotesize{Calculate $d$ and the experimental uncertainty for this single measurement.} & \footnotesize{Students will calculate $d$ and uncertainties using experimental data.} \\ \hline
    
    \footnotesize{Rotate the CD to a few different angles and make measurements of the diffracted beams. Replace with red laser and acquire more data points.} & \footnotesize{Students will make use of different apparatus and adjust them to make measurements and compare results.}  & \footnotesize{Draft a plan that will allow you to find $d$ with higher precision. Identify the physical quantities you will measure, think about how you will measure them, and carry out your plan.} & \footnotesize{Students will reflect on their initial results, design a procedure that will minimize the experimental uncertainties, and carry out their procedure.} \\ \hline
    
    \footnotesize{Use the re-arranged grating equation to plot the data and fit for $d$. Calculate the uncertainty in your estimate.} & \footnotesize{Students will use curve fitting to improve their experimental results and uncertainties.} & \footnotesize{Make use of all your data points to calculate $d$ and your experimental uncertainty. Can you identify any assumptions you have made that might produce systematic uncertainties in your result?} & \footnotesize{Students will evaluate the systematic uncertainties of their experimental design.} \\ \hline

   \end{tabular}
\end{table*}

\section{Methodology}
After redesigning the CD diffraction experiment, we had two different lab manuals at our disposal. We have included them in the supplementary material~\cite{LabManuals}. Although they shared the same experimental goals, and had the same introduction and theory sections, they embodied different levels of student inquiry. The lab manual for the guided-inquiry experiment had a recipe-like procedure with step-by-step instructions. In contrast, the lab manual for the open-inquiry experiment contained a section called ``design activities'' with prompts asking students to design their own activities. Table~\ref{tab:outlineComp} compares the experimental sections and the corresponding learning outcomes for the guided- and open-inquiry experiments.

These manuals were used in the following fashion in the second-year laboratory course in the winter semester of 2017. 
Eight groups of students (in pairs) performed the open-inquiry experiment in the middle of the semester and another five groups (in pairs) performed the guided-inquiry experiment towards the end of the semester. We designed this study this way so that the students who did the open-inquiry experiment did not have access to the guided-inquiry lab manual. Each group was given three hours to complete the experiment. Ethics approval was obtained for the study and student consent was acquired for audio-recording their conversations during the experiment. 

On average, it took 2 hours and 5 minutes for a group to complete the guided-inquiry experiment, while it took 2 hours and 21 minutes for a group to complete the open-inquiry experiment. The audio data files were transcribed using a commercial software called Transcribe~\cite{transcribewebsite} by two senior physics undergraduate research assistants with previous transcribing experiences. The two students in all groups were anonymous and identified as Student A and Student B in the transcripts. The transcribers made sure that they were familiar with the students' voices in their randomly assigned audio files before they started transcribing. Each conversation was stamped with a time corresponding to the start of the audio file. A quality assurance check of the transcripts revealed their high accuracy in reflecting the audio recordings.

This paper focuses on the qualitative and quantitative analysis of the transcripts by addressing the following research questions:
\begin{enumerate}[label=\Alph*., noitemsep,nolistsep] 
\item How do student learning experience and behaviour differ in experiments that allow for either guided or open inquiry? 
\item What level of inquiry better reinforces student experimental design skills? 
\end{enumerate}

Thematic analysis, using the coding procedure described by Corbin and Strauss~\cite{Corbin2015},  was carried out to analyze the audio transcripts using a general inductive approach~\cite{Thomas2006}. General inductive analysis refers to an approach that primarily uses reading of data to derive themes through interpretations of the data made by the researcher. Strategies used in this approach include questioning what core meanings are evident in the text in relation to the research questions. This is achieved by carrying out open, axial, and selective coding steps. Open coding involves identifying 
phrases within text and creating codes for these. The codes are then grouped in the axial coding phase to create categories. Finally, selective coding is completed when the main themes emerging from the categories are created.  The themes addressing the research questions are the findings of the study. Emergent themes have been identified from data of each of the guided- and open-inquiry experiments. These themes, compared to the intended physics laboratory learning outcomes, provide an illustration of what influence the two different experimental experiences have on student learning.

The guided- and open-inquiry experiment transcripts were analyzed separately, that is, the analysis processes for one set of experimental data was kept distinct from the other, in an effort to reduce bias during the coding process. Before analysis began, we kept bracketing notes for awareness of personal biases and research biases. Since the research aims to compare the learning occurring during two experiments, we treated the guided-inquiry transcripts as the control data set, and so this data set was analyzed first before the open-inquiry transcripts. LM, one of the authors, independently completed the open coding of the first guided-inquiry transcript while BC independently completed the open coding of the first 15 minutes of the same transcript. We compared and discussed our codes. LM wrote a description for all her open codes to ensure an accurate understanding of the data and to ensure future transcripts would be approached with a thoroughly developed initial set of codes. BC adopted LM's open codes, continued coding part of the first guided-inquiry transcript, compared with LM's, and discussed the differences until 100\% agreement was reached. This process continued until coding of the first guided-inquiry transcript from the two coders was completely agreed upon. LM then coded the 4 remaining guided-inquiry transcripts. New codes were generated and a few existing codes were modified when necessary. The same coding process was followed for the 8 open-inquiry transcripts.

Axial and selective coding to identify categories and themes for the guided- and then open-inquiry transcripts followed the initial coding phase.

\section{Results}
We report our findings in two sections under qualitative and quantitative results. Qualitative data analysis revealed the major themes associated with student learning and behaviour during the laboratory experience. Quantitative data analysis captured the number of categories, codes, and references in each of the themes in the guided- and open-inquiry experiments. Comparisons between theses themes are presented in the discussion section.
\newcolumntype{L}[1]{>{\raggedright\let\newline\\\arraybackslash\hspace{0pt}}m{#1}}
\begin{table*}[t]
   \caption{Themes and categories that emerged from qualitative coding analysis of the guided-inquiry data set and the open-inquiry data set.}  \vspace{0.2cm}
   \label{tab:themes} \centering
   \begin{tabular}{ L{3.3cm} L{5.8cm} | L{3.3cm} L{3.5cm} }  \hline 
    \multicolumn{2}{c|}{Guided-Inquiry Experiment} & \multicolumn{2}{c}{Open-Inquiry Experiment}\\
    Theme & Category & Theme & Category \\\hline
    \multirow{6}{3.3cm}{Carrying out the experiment }& Constructing apparatus & \multirow{6}{3.3cm}{Experimental process \\and components} & \\
    & Making predictions && \\
    & Measurement and calculation && Process\\
    & Problem solving && Measurement and analysis \\
    & Project management &&\\
    & Visual observation&&\\\hline

    \multirow{5}{3.3cm}{Interpersonal learning} & Needing help & \multirow{5}{3.3cm}{Self and interpersonal experiences} & \\ 
    & Physics language && \multirow{3}{3.5cm}{Attitude\\Interpersonal interactions}\\
    & Peer interaction && \\
    & Student interaction with instructor &&\\
    & Instructor interaction with students &&\\\hline

    & Application && \\ 
    & Confirmation && \\
    & Learning && \\
    & Question asking & \multirow{3}{3.3cm}{Comprehension experience} & \multirow{3}{3.5cm}{Learning\\Question asking\\Sense making}\\
    \multirow{1}{3.3cm}{Sense making} & Sense making  && \\
    & Sense making strategies && \\
    & Sense making of: calculations or measurements, instructions, observations, peer's ideas or suggestions, the task, and physics concepts && \\\hline
    
    \multirow{3}{3.3cm}{The affective experience} & Reactions & \multirow{3}{3.3cm}{Experimental design experience} & \multirow{3}{3.5cm}{Critical thinking\\Design}\\
    & Marvelling && \\
    & Emotions &&\\\hline
   \end{tabular}
\end{table*}

\subsection{Qualitative Analysis Results}
Table~\ref{tab:themes} depicts qualitative coding analysis results. Shown separately are the guided-inquiry data set and the open-inquiry data set with emergent themes and the categories that comprise each theme.

\subsubsection{Guided Inquiry}
Four major themes emerged from the qualitative analysis of the guided-inquiry data: 1) Carrying out the experiment; 2) Interpersonal learning; 3) Sense making; and 4) The affective experience. 

``Carrying out the experiment'' is intended to represent students' experiences of progressing through the steps prescribed by the lab manual until the end is reached. Student behaviours included constructing apparatus, making predictions, making measurements, calculations and visual observations, and engaging in problem solving and project management. We point to the nature of students' progression through steps in the guided-inquiry experiment, which was found to be oriented toward reaching the end of the lab manual. For example, while students took measurements: 
\begin{quote}\vspace*{-0.05in}
    \textit{Alright do one more, I want to get out of here.} 
\end{quote}\vspace*{-0.05in}
In similar spirit, 
\begin{quote}\vspace*{-0.05in}
    \textit{I don't like the [labs] that make you do the lab in the lab.} 
\end{quote}\vspace*{-0.05in}
Other students showed recognition of the value of laboratory time, but remained focused on executing the bare-minimum steps: 
\begin{quote}\vspace*{-0.05in}
\textit{As much as I'd love to get home we should at least make sure our measurements are kind of precise before we [leave].} 
\end{quote}\vspace*{-0.05in}
Progressing from step to step, students continually refer to the lab manual for direction: 
\begin{quote}\vspace*{-0.05in}
\textit{Let's see what we have to do next,} 
\end{quote}\vspace*{-0.05in}
and 
\begin{quote}\vspace*{-0.05in}
\textit{Alright, repeat with the other laser.}
\end{quote}\vspace*{-0.05in}

``Interpersonal learning'' represents students' experiences learning from and with others, including their laboratory partner, other groups, and the instructor or teaching assistant. Experiences related to interpersonal learning included needing help, using physics language, and a host of specific peer interaction modes (e.g., directing, suggesting, informing, cooperation). Student learning from and with others is exemplified in the way that a student's peers and instructors provided a source for their learning. Learning what $n$ is ($n$ is the $n^{\rm th}$ order diffraction and $\lambda$ is the wavelength of the laser light), in this conversation: 
\begin{quote} \vspace*{-0.05in}
A: \textit{n$\lambda$ divided by sine of the $\theta$.}\\
B: \textit{What's n?}\\
A: \textit{n is just our number, like diffraction grating when \dots it's one.}\\
B: \textit{So one $\lambda$?}\\
A: \textit{Yeah.}\\
B: \textit{Oh okay.}
\end{quote}\vspace*{-0.05in}
In another example, reconciling understanding of measurements: 
\begin{quote}\vspace*{-0.05in}
    A: \textit{And then we're measuring $N_1$ and $N_2$, and that's the difference that is going to give us the Y.}\\
    B: \textit{I think you're thinking about measuring L, like the distance to the horizontal. We did that once. But the distance to this point is going to change every time we change the angle right?}
\end{quote}\vspace*{-0.05in}

``Sense making'' represents students' experiences rationalizing, figuring out, or giving meaning to the task at hand. Students made sense of calculations or measurements, instructions, observations, physics concepts, their peer's ideas or suggestions, and of the task they were given. Application of knowledge gained at another time or place occurred as part of the sense making experience, as did behaviours of confirmation, asking questions, and using other sense making strategies. Other sense making strategies used by students in the guided-inquiry experiment included asking the instructor or teaching assistant for assistance, looking at the diagram in the lab manual, or drawing a diagram themselves. Audio recordings from the laboratories include, for example: 
\begin{quote}\vspace*{-0.05in}
\textit{I feel like I'd prefer to draw this \dots I just think better on paper I guess,} \textit{I just have to look at the picture really quick.} 
\end{quote}\vspace*{-0.05in}

``The affective experience'' represents students' feelings, emotions, moods, and attitudes related to the laboratory experience. Specifically, this experience includes reactions, marveling, positive emotions such as confidence, determination, excited, hopeful, and having fun, and negative emotions such as careless, confusion, frustration, disappointment, sarcasm, self-doubt, and stress. Students' expressions of confusion, for example, relate to their understanding of variables: 
\begin{quote}\vspace*{-0.05in}
  \textit{I understand what they're asking us to do with this equation but I don't understand how, how our n value is supposed to change,}  
\end{quote}\vspace*{-0.05in}
and 
\begin{quote}\vspace*{-0.05in}
\textit{I'm still confused about which one is the, like, zero order.}  
\end{quote} \vspace*{-0.05in}
On the opposite side of the affective experience, students felt confident in following the procedure:
\begin{quote}\vspace*{-0.05in}
\textit{Well I'm pretty confident in our procedure so I guess we can just take these measurements pretty quick.} 
\end{quote}\vspace*{-0.05in}
As demonstrated by the range of aspects that were part of the students' affective experience, students completing the guided-inquiry experiment communicated a breadth of emotions.

\subsubsection{Open Inquiry}
Four major themes emerged from the qualitative analysis of the open-inquiry data: 1) Experimental process and components; 2) Self and interpersonal experiences; 3) Comprehension experience; and 4) Experimental design experience.

``Experimental process and components'' represents students' experiences working in a non-stepwise process of experimentation in the open-inquiry experiment. The nature of this experimental process and its components for students was task-oriented, meaning that they were given the task to achieve accurate measurements and their learning experience and behaviour were oriented toward accomplishing the task. The nature of this process in the open-inquiry experiment did not resemble sequential completion of steps, or an end-oriented experience. The process included learning experience and behaviour related to measurement and analysis and the overall process, which was categorized by the following: configuring laboratory station, observations, being off task, recording activity, referring to resources, safety, and troubleshooting. With this nature of experimental process and components, students accomplished the task they were given in the open-inquiry experiment. Students were interpreting their measurements: 
\begin{quote}\vspace*{-0.05in}
\textit{I feel like this is going to be worse of a reading or even greater experimental error on our part,} 
\end{quote}\vspace*{-0.05in}
and
\begin{quote}\vspace*{-0.05in}
\textit{Got a perfect range of values.} 
\end{quote}\vspace*{-0.05in}
Students also considered how to graph the data they gathered: 
\begin{quote}\vspace*{-0.05in}
\textit{I'm trying to think of all the things we could plot,} 
\end{quote}\vspace*{-0.05in}
and 
\begin{quote}\vspace*{-0.05in}
    \textit{I was trying to calibrate that, some sort of relationship where d was the slope of something.}
\end{quote}\vspace*{-0.05in}

``Self and interpersonal experiences'' are those that relate to students' own attitudes and their interactions with others. The aspects of their experience that relate to attitude include identity, interest, marvelling, and wanting to finish the experiment. The aspects of their experience that relate to interpersonal experiences include interaction with other groups, the instructor or teaching assistant, and their peers. Students had a certain positionality or frame of attitude in the open-inquiry experiment; for example, only one student remarked on wanting to finish the experiment, while other students marvelled, making comments such as: 
\begin{quote}\vspace*{-0.05in}
\textit{Whoa, what are you guys doing? That looks so cool.} 
\end{quote}\vspace*{-0.05in}
Another student, reflecting on their experience of the experiment, remarked that their group were
\begin{quote}\vspace*{-0.05in}
\textit{such scientists.} 
\end{quote}\vspace*{-0.05in}
In terms of interpersonal experiences, students' interaction with the instructor or teaching assistant varied from asking questions, asking for help, and explaining their design. When the instructor approached the students, interactions included explaining concepts or the purpose of the experiment, checking in, answering questions, encouragement, guidance, and prompting. Interaction with other groups involved comparison or gaining understanding from them. Peer interaction within the laboratory groups ranged widely and included these behaviours: clarification, informing, dismissing, suggesting, showing peer, planning, reassurance, and others related to language and cooperation.

``Comprehension experience'' represents students' experience of coming to understand the task at hand that is related to students' comprehension or understanding that occurred via learning, question asking, and sense making. An example of a student's learning experience in the open-inquiry experiment is: 
\begin{quote}\vspace*{-0.05in}
\textit{Oh wow that was so \dots now I understand what's going on. Took me, like, the first hour to figure out what we were, like, the theory behind all of this stuff. Once I understood \dots} 
\end{quote}\vspace*{-0.05in}
Such learning experiences included reflection on the experiment design or process. For example,
\begin{quote}\vspace*{-0.05in}
    A: \textit{We can just start trying things and then see, but I feel like \dots}\\
    B: \textit{I feel like the original method we did was pretty good.}
\end{quote}\vspace*{-0.05in}
Questions that students asked ranged in nature, from calculation and measurement questions to design questions and next steps questions. Other types included questions related to data collection, using equipment, observations, and understanding. Understanding question examples include: 
\begin{quote}\vspace*{-0.05in}
\textit{How do lasers work?} \\
\textit{How does this give us the average of d?} 
\end{quote}\vspace*{-0.05in}
and 
\begin{quote}\vspace*{-0.05in}
\textit{Is the \dots how is the \dots line spacing directly proportional to the distance between the diffraction lines?}
\end{quote} \vspace*{-0.05in}
Design question examples include: 
\begin{quote}\vspace*{-0.05in}
\textit{Okay, so what can we change?} \\
\textit{Why does it even have to be level, what difference does it make?} 
\end{quote}\vspace*{-0.05in}
and 
\begin{quote}\vspace*{-0.05in}
\textit{How did we know that this is going to hit here and then go directly to the center, like it hit the center of this CD?} 
\end{quote}\vspace*{-0.05in}
Sense making experiences included topics similar to those that students asked questions about, such as calculations, data, measurements, observations, and instructions. However, students also made sense of diagrams, equations, tasks, and the apparatus.

``Experimental design experience'' represents the students' intentional consideration, creation, and execution of plans in the open-inquiry experiment. These experiences were grouped into two categories: critical thinking and design. Critical thinking behaviours are exemplified by comments such as: 
\begin{quote}\vspace*{-0.05in}
\textit{Okay, so. Rotating the CD, how does that affect \dots that affects $\theta$ too, right? So let's try this out,} 
\end{quote}\vspace*{-0.05in}
and
\begin{quote}\vspace*{-0.05in}
\textit{If you look at this, the laser light is pointing almost right back into the laser itself, right? So that would indicate, if this is already level, then that should be.} 
\end{quote}\vspace*{-0.05in}
Design experience and behaviour were further grouped into two categories: constructing apparatus and designing experiment. Constructing apparatus involved making assumptions, being resourceful, concern, designing the apparatus, making adjustments, trial, and using equipment. Designing experiment involved a variety of design-oriented learning experience and behaviour, including foreseeing issues, decision-making, changing directions, improvement ideas, making predictions, interpreting, and testing these, and identifying tasks, a goal or purpose, and known and unknown information. The following quotation broadly exemplifies the open-inquiry student experience of designing the experiment: 
\begin{quote}\vspace*{-0.05in}
\textit{So I'm thinking we can \dots do several different trials by using different distance from the mirror to the wall.}
\end{quote}\vspace*{-0.05in}

\subsection{Quantitative Analysis Results}
We quantified the codes, categories, and themes for each of the guided- and open-inquiry data sets to broadly represent the analysis results. Table~\ref{tab:GuidedNumOfCodes} and \ref{tab:OpenNumOfCodes} show the number of categories, codes, and references in each of the themes in the guided- and open-inquiry experiments. These numbers may provide insight about the differences in the nature of students' learning experiences in the two different experiments. Student experiences in the guided-inquiry experiment span almost three times the number of categories when compared to those in the open-inquiry experiment. Conversely, when examining by the smallest unit, the number of codes, student experiences in the open-inquiry experiment are represented by more than ten percent more codes when compared to those in the guided-inquiry experiment. In the discussion section we explore how these initial differences may suggest a more concentrated, deep-level experience in the open-inquiry experiment compared to that in the guided-inquiry experiment.
\begin{table}[ht!]
   \caption{The number of categories, codes, and references in each of the themes in the guided-inquiry experiment. Note that there are 5 guided-inquiry data sets from the 5 groups of students who did the guided-inquiry experiment.}  \vspace{0.15cm}
   \label{tab:GuidedNumOfCodes} \centering
   \begin{tabular}{ L{4.3cm} L{1.4cm} L{0.8cm} L{1.5cm} }   \hline
    Guided-Inquiry Theme & No. Categories & No. Codes & No. References \\  \hline 
    Carrying out the experiment & 6 & 39 & 890\\
    Interpersonal learning & 5 & 23 & 553\\
    Sense making & 12 & 21 & 395\\
    The affective experience & 3 & 16 & 130\\    \hline
    Total for guided inquiry & 26 & 99 & 1968\\   \hline
    \end{tabular}
\end{table}
\begin{table}[ht!]
   \caption{The number of categories, codes, and references in each of the themes in the open-inquiry experiment. Note that there are 8 open-inquiry data sets from the 8 groups of students who did the open-inquiry experiment.}  \vspace{0.15cm}
   \label{tab:OpenNumOfCodes} \centering
   \begin{tabular}{ L{4.3cm} L{1.4cm} L{0.8cm} L{1.5cm} }   \hline
    Open-Inquiry Theme & No. Categories & No. Codes & No. References \\  \hline
    Experimental process and components & 2 & 32 & 708\\
    Self and interpersonal experiences & 2 & 31 & 366\\
    Comprehension experience & 3 & 26 & 353\\
    Experimental design experience & 2 & 23 & 444\\  \hline
    Total for open inquiry & 9 & 112 & 1871\\  \hline
    \end{tabular}
\end{table}

\begin{figure}[ht!]
\includegraphics[width=1.0\linewidth]{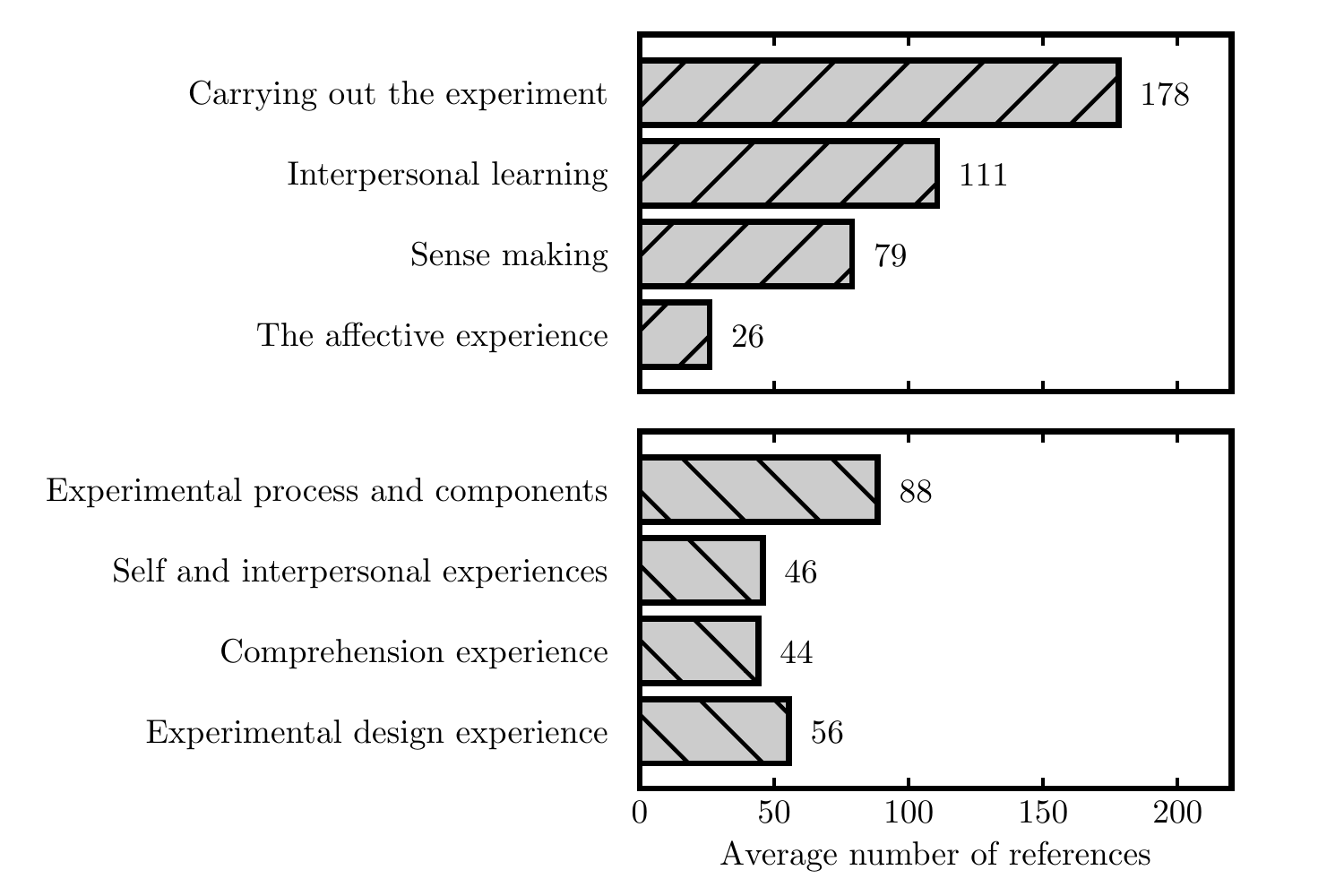}
\caption{Histograms depicting the frequency of experience across themes in each experiment. Shown are the average number of references in each theme in the guided-inquiry experiment (top) and in the open-inquiry experiment (bottom). The frequency is calculated by dividing the total number of references in each theme by the number of the corresponding data sets in the guided-inquiry (5 data sets) or open-inquiry experiment (8 data sets). \label{fig:freqHist}}\vspace*{-0.15in}
\end{figure}

As for the remaining quantitative analysis results, we have created plots to depict the frequencies of themes, categories, and codes for both the guided-inquiry and open-inquiry experiments. Figure~\ref{fig:freqHist} shows the frequency of each theme based on the average number of references (coded segments of raw text) contributing to it, calculated by dividing the total number of references across all the codes in the theme by the number of the corresponding data sets in the guided-inquiry or open-inquiry experiment. 

\begin{figure*}[pht!]
\centering \hspace*{-1cm}
\includegraphics[width=2.3\columnwidth]{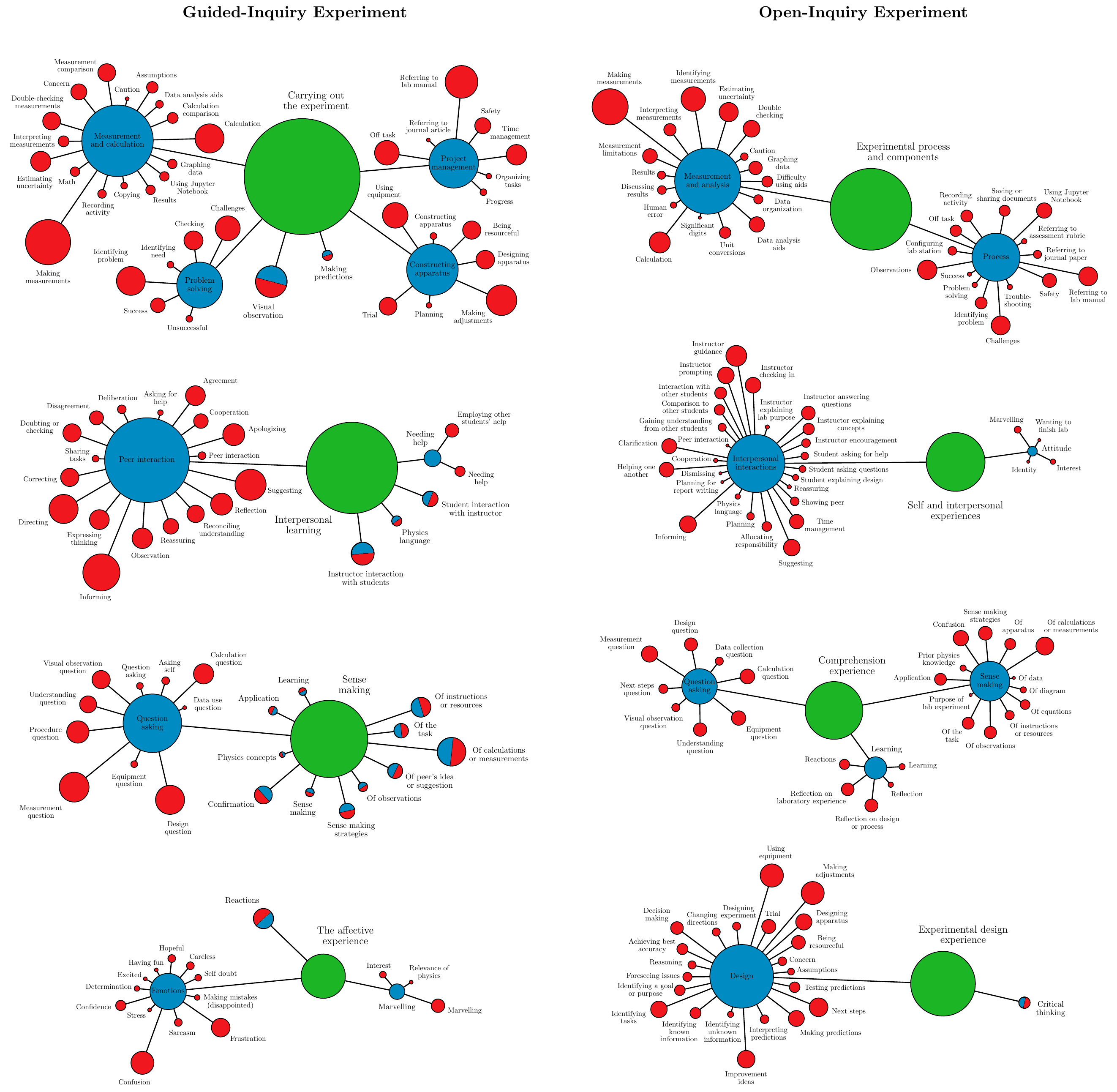}
\caption{This figure represents the complete picture of the our qualitative and quantitative analysis results. All categories and codes in each theme are shown for the guided-inquiry experiment (left) and for the open-inquiry experiment (right). The area of each node scales with the frequency of the theme (green), category (blue), and code (red). When a circle is seen half filled with blue and half filled with red, this is an indication that the circle represents a term that is both a category and code. An example of this is ``Physics language,'' seen in the ``Interpersonal learning'' theme in the guided-inquiry experiment on the left.}\label{fig:allCodesBubblePlot}
\end{figure*}
Figure~\ref{fig:allCodesBubblePlot} shows in more detail all the categories and codes in each theme with the frequency of the corresponding theme, category, and code represented by the area of the circle. To interpret the relative frequency at the level of codes, for example, readers can compare the areas of red circles. In an effort to ensure the figure's readability, we provide an example: ``Asking for help,'' a code within the category ``Peer interaction'' and the theme ``Interpersonal learning,'' is the smallest red circle among its fellow red circles. This means that ``Asking for help'' was least frequently coded for during analysis in comparison to other codes in this category. Continuing with the example, ``Deliberation,'' seen to the left of ``Asking for help,'' was more frequently coded for.

By depicting the qualitative analysis in terms of frequency of codes, we are able to visually represent the student learning experience and behaviour in each of the guided- and open-inquiry experiments. In the guided-inquiry experiment, student learning experience and behaviour were most frequently coded as ``Carrying out the experiment,'' within which students most frequently engaged in measurement and calculation, and to lesser frequencies in constructing apparatus, project management, and problem solving.  The second highest frequency theme is ``Interpersonal learning,'' with the highest frequency category being peer interaction, followed by other similarly lower frequency categories such as instructor interaction with students, student interaction with instructor, needing help, and physics language. The third highest frequency theme is ``Sense making,'' with its highest frequency category question asking, and lower frequency categories including sense making of calculations or measurements, of instructions, and using sense making strategies. The least frequent theme in the guided-inquiry experiment is ``The affective experience,'' with the high-frequency category being emotion (including a variety of types such as confusion, frustration, and confidence) and the lower frequency category being marvelling.

In the open-inquiry experiment, student learning experience and behaviour were most frequently coded as ``Experimental process and components,'' within which students most frequently engaged in measurement and analysis and less frequently in process, which includes frequent experience and behaviour such as referring to the lab manual, challenges, and observations. The second highest frequency theme is ``Experimental design experience,'' with the highest frequency category being design. Design includes a variety of experience and behaviour, most frequent of which include using equipment, making adjustments, next steps, and improvement ideas. The only other category in this theme is critical thinking. The third highest frequency theme is ``Self and interpersonal interactions.'' The higher frequency category in this theme is interpersonal interactions, which includes numerous related behaviours such as instructor interaction with students, interaction with other groups, and informing (peers). The least frequent theme in the open-inquiry experiment is ``Comprehension experience,'' within which students most frequently engaged with sense making, and to lesser frequencies in question asking and learning. 

\section{Discussion}
Overall, the guided- and open-inquiry experiments each revealed four different themes, which allowed us to understand the student learning behaviours in each of these experiences as unique. The extent to which and how the nature of the experiences differed was challenging to capture. In an attempt to illuminate the differences of student learning experience and behaviour between the two forms of inquiry and the benefits to students' learning in the open-inquiry experiment, we discuss how the themes compare between the experiments, including discussions about common and unique codes in each experiment. 

\subsection{Comparing Number of Codes, Categories and Themes}
The number of codes, categories, and themes resulting from analysis of each experiment suggest a difference in the nature of students' learning experience and behaviour in the two different experiments. As reported earlier, the guided- and open-inquiry experiments both yielded 4 themes. The guided-inquiry experiment yielded 26 categories and 99 codes, whereas the open-inquiry experiment yielded 9 categories and 112 codes. What we observe by comparing the number of codes and categories in the open-inquiry experiment is that a high number of codes (112) reduced to a low number of categories (9) shows high instances of experience and behaviour occurring at a low variety. What we observe by comparing the number of codes and categories in the guided-inquiry experiment is that a high number of codes (99) reduced to a medium number of categories (26) shows high instances of experience and behaviour occurring at a high variety. Based on these results we suggest that the open-inquiry experiment allowed students to have more  concentrated, deep-level experiences compared to those in the guided-inquiry experiment.

To provide an example that may illustrate our assertion, we compare the guided-inquiry theme ``Carrying out the experiment'' and the open-inquiry theme ``Experimental process and components.'' Within the former, there are 6 categories (in descending frequency: measurement and analysis, constructing apparatus, project management, problem solving, visual observation, and making predictions) that describe how students carried out the experiment. These categories further divide into 39 codes. Within the open-inquiry theme ``Experimental process and components,'' there are two categories (in descending frequency: measurement and analysis, and process) that describe students' experimental process. These categories further divide into 32 codes. The two themes contain similar number of codes, although in the open-inquiry experiment the students' learning experience and behaviour could be grouped into just two categories, only one-third of the number of categories in the guided-inquiry experiment when comparing these themes. Therefore, we assert that students had a different experience in the open-inquiry experiment than in the guided-inquiry experiment; their inquiry is more concentrated in focus, which we attribute to the open level of inquiry.

\subsection{Frequency Observations}
To further our above assertion and suggest that the open-inquiry experiment is more supportive of students' experimental design skills than the guided-inquiry experiment, we describe observations made of the frequency of themes (see Figure~\ref{fig:freqHist}). Student's experiences in the open-inquiry experiment are most frequently experimental process- and design-related, whereas in the guided-inquiry experiment, student's experiences are most frequently related to carrying out the experiment and interpersonal learning. In addition, the themes aligning more closely with the AAPT's guidelines for designing experiments are the open-inquiry experiment themes. It suggests that the open level of inquiry provides students with a design skill-reinforcing experience more so than the guided level of inquiry. Interestingly, the average total references for the open-inquiry experiment (234) is lower than that for the guided-inquiry experiment (394), which may suggest students simply talked less and engaged in more thinking or doing. While it was beyond the scope of this paper, we suggest these findings of relative quietness in the open-inquiry experiment could support students' exploration, engagement, and the experience of design.

We also compared student's experiences in each experiment by the least frequent themes. The single-lowest frequency theme in the open-inquiry experiment was ``Comprehension experience,'' which is only slightly less frequent than the second-lowest frequency theme, ``Self and interpersonal experience''; however, the lowest frequency theme in the guided-inquiry experiment is ``The affective experience,'' which is less than half as frequent as the second-lowest theme, ``Sense making.'' These findings suggest a more balanced experience across themes in the open-inquiry experiment than in the guided-inquiry experiment. In terms of helping students meet learning outcomes intended to develop their design skills, we find a more balanced laboratory experience to be favourable because not one learning experience or behaviour is favoured too highly over other supporting behaviours. As is the case in the guided-inquiry experiment, the frequency of students' experiences was concentrated on ``Carrying out the experiment'' while the frequency of other experience and behaviour suffered in comparison.

The idea that certain behaviours or learning experiences may happen in support of or at the expense of others is an interesting topic. We have observed ``The affective experience'' as part of the guided-inquiry experience, albeit the least frequent theme, as a possible hindrance to students' experiencing other themes more frequently. Because of the nature of the affective experience which is unique to the guided-inquiry experiment, with students' two most frequent emotions being confusion and frustration, it is possible that the affective experience hindered students from having an experience more frequently design-oriented (or otherwise). The results relating to theme frequency as they are, however, do suggest that students' experiences in the open-inquiry experiment are more frequently design-oriented compared to the guided-inquiry experiment and therefore we advise that the open level of inquiry better reinforces students' design skills. 

\subsection{Common Codes Across Themes}
\begin{table*}[pht!]
   \caption{A list of the 52 common codes that emerged from qualitative coding analysis of the guided-inquiry data set and the open-inquiry data set. We list the 4 themes from the guided-inquiry experiment first, followed by the 4 themes from the open-inquiry experiment. The numbers shown in the table are how many times each code is referenced in the corresponding experiment and how many files each code is referenced (shown in parentheses). Note that the total number of guided- and open-inquiry files is 5 and 8, respectively.}  \vspace*{-0.2cm}
   \label{tab:commoncodes} \centering
  \begin{tabular}{l x{1.3cm} x{1.3cm} x{1.3cm} x{1.3cm} | x{1.3cm} x{1.3cm} x{1.3cm} x{1.3cm}}
 Code
 & \rot[45]{\parbox{2cm}{\raggedright Carrying out the experiment}}
 & \rot[45]{\parbox{1.8cm}{\raggedright Interpersonal learning}}
 & \rot[45]{\parbox{1.8cm}{\raggedright Sense making}}
 & \multicolumn{1}{c}{\rot[45]{\parbox{2cm}{\raggedright The affective experience}}}
 & \rot[45]{\parbox{3cm}{\raggedright Experimental process and components}}
 & \rot[45]{\parbox{3cm}{\raggedright Self and interpersonal experiences}}
 & \rot[45]{\parbox{2.5cm}{\raggedright Comprehension experience}}
 & \rot[45]{\parbox{2.5cm}{\raggedright Experimental design experience}}
  \tabularnewline
    \midrule
    Being resourceful & 22 (5) & - & - & - & - & - & - & 20 (7) \\
    Designing apparatus & 23 (5) & - & - & - & - & - & - & 28 (6) \\
    Making adjustments & 62 (5) & - & - & - & - & - & - & 56 (8) \\
    Planning & 2 (2) & - & - & - & - & 6 (4) & - & - \\
    Trial & 21 (5) & - & - & - & - & - & - & 21 (7) \\
    Using equipment & 43 (5) & - & - & - & - & - & - & 57 (8) \\
    Making predictions & 7 (4) & - & - & - & - & - & - & 27 (7) \\
    Calculation & 56 (4) & - & - & - & 47 (8) & - & - & - \\
    Data analysis aids & 4 (2) & - & - & - & 25 (4) & - & - & - \\
    Graphing data & 6 (2) & - & - & - & 19 (5) & - & - & - \\
    Assumptions & 9 (4) & - & - & - & - & - & - & 5 (5) \\
    Caution & 1 (1) & - & - & - & 6 (5) & - & - & - \\
    Concern & 17 (4) & - & - & - & - & - & - & 8 (6) \\
    Double checking & 21 (4) & - & - & - & 30 (7) & - & - & - \\
    Estimating uncertainty & 27 (3) & - & - & - & 39 (7) & - & - & - \\
    Interpreting measurements & 8 (3) & - & - & - & 16 (7) & - & - & - \\
    Making measurements & 135 (5) & - & - & - & 138 (8) & - & - & - \\
    Recording activity & 5 (1) & - & - & - & 15 (6) & - & - & - \\
    Using Jupyter Notebook & 6 (2) & - & - & - & 25 (5) & - & - & - \\
    Results & 6 (3) & - & - & - & 7 (2) & - & - & - \\
    Challenges & 41 (5) & - & - & - & 38 (7) & - & - & - \\
    Identifying problem & 55 (5) & - & - & - & 15 (6) & - & - & - \\
    Success & 14 (4) & - & - & - & 2 (2) & - & - & - \\
    Off task & 40 (5) & - & - & - & 14 (5) & - & - & - \\
    Referring to journal article & 1 (1) & - & - & - & 7 (3) & - & - & - \\
    Referring to lab manual & 71 (5) & - & - & - & 38 (7) & - & - & - \\
    Safety & 17 (5) & - & - & - & 21 (7) & - & - & - \\
    Time management & 28 (4) & - & - & - & - & 22 (6) & - & - \\
    Physics language & - & 7 (3) & - & - & - & 3 (3) & - & - \\
    Peer interaction & - & 4 (1) & - & - & - & 1 (1) & - & - \\
    Cooperation & - & 14 (4) & - & - & - & 2 (2) & - & - \\
    Informing & - & 91 (5) & - & - & - & 30 (6) & - & - \\
    Reassuring & - & 16 (4) & - & - & - & 2 (2) & - & - \\
    Reflection & - & 32 (5) & - & - & - & - & 3 (2) & - \\
    Suggesting & - & 64 (5) & - & - & - & 29 (6) & - & - \\
    Application & - & - & 5 (2) & - & - & - & 15 (6) & - \\
    Learning & - & - & 4 (3) & - & - & - & 4 (2) & - \\
    Calculation question & - & - & 27 (4) & - & - & - & 23 (6) & - \\
    Design question & - & - & 56 (5) & - & - & - & 18 (7) & - \\
    Equipment question & - & - & 3 (1) & - & - & - & 22 (8) & - \\
    Measurement question & - & - & 59 (5) & - & - & - & 28 (7) & - \\
    Understanding question & - & - & 19 (5) & - & - & - & 19 (7) & - \\
    Visual observation question & - & - & 22 (5) & - & - & - & 7 (5) & - \\
    Sense making strategies & - & - & 16 (5) & - & - & - & 20 (7) & - \\
    Of calculations or measurements & - & - & 54 (5) & - & - & - & 34 (8) & - \\
    Of instructions or resources & - & - & 26 (5) & - & - & - & 9 (5) & - \\
    Of observations & - & - & 6 (3) & - & - & - & 16 (7) & - \\
    Of the task & - & - & 14 (5) & - & - & - & 15 (7) & - \\
    Reactions & - & - & - & 27 (5) & - & - & 11 (6) & - \\
    Marvelling & - & - & - & 12 (3) & - & 5 (4) & - & - \\
    Interest & - & - & - & 3 (2) & - & 3 (3) & - & - \\
    Confusion & - & - & - & 35 (5) & - & - & 25 (8) & - \\
    \bottomrule
\end{tabular}
\end{table*}
We examine the common codes that emerged in both guided- and open-inquiry experiments in this section, i.e., those that represent a specific learning experience and/or behaviour that occurred for students in both data sets. We explore interpretations of such results as a means to address how student learning experience and behaviour differ in guided- and open-inquiry experiments; this is our first research question. The common codes emerged in both experiments as expected since a) the students did the CD diffraction experiment with the same set of provided apparatus; b) there are common learning outcomes shared within the two levels of inquiry for this experiment as seen in table~\ref{tab:outlineComp}. We list in detail these common codes in table~\ref{tab:commoncodes} with their frequencies of being referenced in the corresponding themes in the two levels of inquiry. 

Most of the common codes emerged in the ``Carry out the experiment'' theme in the guided-inquiry experiment, which is also the theme that has the largest number of codes and references. These common codes are either in the ``Experimental process and components'' or the ``Experimental design experience'' theme in the open-inquiry experiment, except that the ``Planning'' code is in the ``Self and interpersonal experience'' theme. Some of these common codes have similar frequencies in the two experiments while many have large differences in frequencies. For example, students made similar amount of efforts in ``Being resourceful,'' ``Designing apparatus,'' making ``Trial,'' ``Using equipment,'' and ``Double checking'' measurements in the two experiments, and often talked about ``Safety.'' Notably, students did a lot more ``Making adjustments,'' ``Calculation,'' ``Estimating uncertainty,'' ``Making measurements,'' ``Identifying problem,'' and ''Refering to lab manual'' throughout the guided-inquiry experiment compared to the open-inquiry experiment. They also encountered more ``Challenges'' and went ``Off task'' a lot more often. On the other hand, students in the open-inquiry experiment did more ``Making predictions'' and ``Using Jupyter Notebook,'' and used ``Data analysis aids'' more often than those in the guided-inquiry experiment.

Some common codes emerged both in the ``Interpersonal learning'' theme in guided inquiry and in the ``Self and interpersonal experiences'' theme in open inquiry. Many of the student conversations in guided inquiry fell into the ``Peer interaction'' category and the students were most often ``Informing'' and ``Suggesting'' ideas to each other. The natures of the conversations in open inquiry were quite different and were often more design based. Therefore many of the conversations moved away from simply ``Informing'' or ``Suggesting'' and fell into the unique codes in the ``Design'' category in the ``Experimental design experience'' theme, discussed in the next section.

We also saw that students in guided inquiry had more questions (a total of 227 references in 5 data sets) than those in open inquiry (a total of 133 references in 8 data sets). The common codes in the ``Sense making'' theme in guided inquiry that are more frequent than those in the ``Comprehension experience'' theme in open inquiry are: ``Calculation question,'' ``Design question,'' ``Measurement question,'' ``Visual observation question,'' ``Sense making of calculations or measurements,'' and ``Sense making of instructions or resources.'' On the other hand, the ``Equipment question'' code was much more frequent in open inquiry. 

Lastly, the ``Reactions'' and ``Confusion'' codes emerged much more frequently in the ``The affective experience'' theme in guided inquiry than in the ``Comprehension experience'' theme in open inquiry. For example, there were a total of 35 references of ``Confusion'' in the 5 guided-inquiry data sets, and 25 references in the 8 open-inquiry data sets, indicating that students on average experienced confusion less frequently in the open-inquiry experiment.

\subsection{Unique Codes Across Themes}
Many unique codes emerged in either the guided- or open-inquiry experiment that highlight the differences in these two experiments. We have listed the 47 codes that are unique in the guided-inquiry experiment and the 60 codes that are unique in the open-inquiry experiment in table~\ref{tab:Unique1} and table~\ref{tab:Unique2}.
\newcolumntype{L}[1]{>{\raggedright\let\newline\\\arraybackslash\hspace{0pt}}m{#1}}
\begin{table*}[t]
   \caption{Open codes that are unique in the ``Carrying out the experiment'' and ``Interpersonal learning'' themes in the guided-inquiry data set in comparison to the ``Experimental process and components'' and ``Self and interpersonal experiences'' themes correspondingly in the open-inquiry data set. Note that there are a total of 5 guided-inquiry files and 8 open-inquiry files.}  \vspace{0.2cm}
   \label{tab:Unique1} \centering
   \begin{tabular}{ | L{0.3cm} | L{2.8cm} L{3.4cm} L{1.5cm} | L{0.3cm} | L{2.4cm} L{3.8cm} L{1.5cm} | }  \hline 
    \multicolumn{4}{|c|}{Guided-Inquiry Experiment} & \multicolumn{4}{c|}{Open-Inquiry Experiment}\\
     & Category & Code & References (Files) & & Category & Code & References (Files)  \\\hline
    \multirow{14}{0.3cm}{\rotatebox{90}{Carrying out the experiment}} & Constructing apparatus & Constructing apparatus & 3 (3) & \multirow{14}{0.3cm}{\rotatebox{90}{Experimental process and components}} & \multirow{6}{2.8cm}{Process} & Configuring lab station & 8 (6)\\\cline{2-4}
    & \multirow{4}{2.8cm}{Measurement and Calculation} & Calculation comparison & 8 (3) &  &  & Observations & 40 (8)\\
    &  &  Math & 6 (3) &  &  & Saving or sharing documents & 13 (7)\\
    &  & Measurement comparison & 21 (3) &  &  & Referring to assessment & 3 (2) \\
    &  & Coping & 3 (2) &  &  & Troubleshooting & 3 (3)\\\cline{2-4}
    & \multirow{3}{2.8cm}{Problem solving} & Checking & 24 (4) &  &  & Problem solving & 3 (2)\\\cline{6-8}
    &  &  Identifying need & 3 (2) &  & \multirow{8}{2.4cm}{Measurement and analysis} & Significant digits & 1 (1)\\
    &  & Unsuccessful & 3 (2) &  &  & Unit conversions & 14 (6)\\\cline{2-4}
    & \multirow{2}{2.8cm}{Project management} & Organizing tasks & 2 (1) &  &  & Data organization & 9 (3)\\
    &  &  Progress & 3 (2) &  &  & Difficulty using aids & 13 (5) \\\cline{2-4}
    &  Visual observation & Visual observation & 66 (5) &  &  & Identifying measurements & 64 (8) \\
    &  &  &  &  &  & Measurement limitations & 23 (7) \\
    &  &  &  &  &  & Discussing results & 8 (5) \\
    &  &  &  &  &  & Human error & 4 (3)\\
    \hline\hline
    \multirow{21}{0.3cm}{\rotatebox{90}{Interpersonal learning}} & \multirow{2}{2.8cm}{Needing help} & Needing help & 7 (4) & \multirow{21}{0.3cm}{\rotatebox{90}{Self and interpersonal experiences}} & \multirow{2}{2.4cm}{Attitude} & Identity & 1 (1)\\
    &  & Employing other students' help & 12 (3) &  &  & Wanting to finish the lab & 1 (1)\\\cline{2-4}\cline{6-8}
    & \multirow{12}{2.8cm}{Peer interaction} & Apologizing & 32 (4) & & \multirow{19}{2.4cm}{Interpersonal interactions} & Instructor answering questions & 20 (6)\\
    &  & Agreement & 26 (5) &  &  & Instructor checking in & 26 (8) \\
    &  & Asking for help & 2 (1) &  &  & Instructor encouragement & 8 (4) \\
    &  & Deliberation & 5 (2) &  &  & Instructor explaining concepts & 14 (7) \\
    &  & Disagreement & 13 (3) &  &  & Instructor explaining the purpose of the lab & 2 (1) \\
    &  & Doubting or checking & 22 (4) &  &  & Instructor guidance & 46 (8) \\
    &  & Sharing tasks & 3 (2) &  &  & Instructor prompting & 29 (8) \\
    &  & Correcting & 22 (5) &  &  & Interaction with other students & 15 (5) \\
    &  & Directing & 56 (5) &  &  & Comparison to other students & 12 (5) \\
    &  & Expressing thinking & 26 (5) &  &  & Gaining understanding from other students & 7 (3) \\
    &  & Observation & 28 (5) &  &  & Clarification & 24 (8) \\
    &  & Reconciling understanding & 20 (5) &  &  & Helping one another & 23 (8) \\\cline{2-4}
    & Student interaction with instructor & Student interaction with instructor & 16 (4) &  &  & Dismissing & 1 (1) \\\cline{2-4}
    & Instructor interaction with students & Instructor interaction with students & 35 (5) &  &  & Allocating responsibility & 10 (5) \\
    &  &  &  &  &  & Planning for report writing & 1 (1) \\
    &  &  &  &  &  & Showing peer & 8 (5) \\
    &  &  &  &  &  & Student asking for help & 6 (5) \\
    &  &  &  &  &  & Student asking questions & 5 (4) \\
    &  &  &  &  &  & Student explaining design & 4 (2) \\
    \hline
    \end{tabular}
\end{table*}

\newcolumntype{L}[1]{>{\raggedright\let\newline\\\arraybackslash\hspace{0pt}}m{#1}}
\begin{table*}[t]
   \caption{Open codes that are unique in the ``Sense making'' and ``The affective experience'' themes in the guided-inquiry data set in comparison to the ``Comprehension experience'' and ``Experimental design experience'' themes correspondingly in the open-inquiry data set. Note that there are a total of 5 guided-inquiry files and 8 open-inquiry files.}  \vspace{0.2cm}
   \label{tab:Unique2} \centering
   \begin{tabular}{| L{0.3cm} | L{2.8cm} L{3.4cm} L{1.5cm} | L{0.3cm} | L{2.8cm} L{3.4cm} L{1.5cm} | }  \hline 
    \multicolumn{4}{|c|}{Guided-Inquiry Experiment} & \multicolumn{4}{c|}{Open-Inquiry Experiment}\\
     & Category & Code & References (Files) & & Category & Code & References (Files)  \\\hline
    \multirow{8}{0.3cm}{\rotatebox{90}{Sense making}} & Confirmation & Confirmation & 21 (3) & \multirow{10}{0.3cm}{\rotatebox{90}{Comprehension experience}} & \multirow{2}{2.8cm}{Learning} & Reflection on design or process & 18 (6)\\\cline{2-4}
    & \multirow{4}{2.8cm}{Question asking} & Question asking & 3 (3) &  &  & Reflection on laboratory experience & 17 (4)\\\cline{6-8}
    &  & Asking self & 4 (3) &  & \multirow{2}{2.8cm}{Question asking} & Data collection question & 7 (4)\\
    &  & Data use question & 1 (1) &  &  & Next steps question & 9 (3)\\\cline{6-8}
    &  & Procedure question & 33 (5) &  & \multirow{6}{2.8cm}{Sense making} & Prior physics knowledge & 4 (2)\\\cline{2-4}
    & Sense making & Sense making & 5 (2) &  &  & Of apparatus & 13 (6) \\\cline{2-4}
    & Of peer's idea or suggestion & Of peer's idea or suggestion & 15 (5) &  &  & Of data & 1 (1) \\\cline{2-4}
    & Physics concepts & Physics concepts & 2 (1) &  &  & Of diagram & 4 (3)\\
    &  &  &  &  &  & Of equations & 10 (4) \\
    &  &  &  &  &  & Purpose of lab experiment & 1 (1)\\
    \hline
    \hline
    \multirow{12}{0.3cm}{\rotatebox{90}{The affective experience}} & Marvelling & Relevance of physics & 1 (1) & \multirow{15}{0.3cm}{\rotatebox{90}{Experimental design experience}} & Critical thinking & Critical thinking & 14 (6)\\\cline{2-4}\cline{6-8}
    & \multirow{11}{2.8cm}{Emotions} & Careless & 4 (3) &  & \multirow{14}{2.8cm}{Design} & Designing experiment & 7 (5)\\
    &  & Frustration & 23 (5) &  &  & Changing directions & 7 (5) \\
    &  & Making mistakes (disappointed) & 2 (2) &  &  & Decision making & 17 (5) \\
    &  & Sarcasm & 4 (2) &  &  & Achieving best accuracy & 13 (5) \\
    &  & Self doubt & 3 (3) &  &  & Reasoning & 7 (5) \\
    &  & Stress & 1 (1) &  &  & Foreseeing issues & 9 (6) \\
    &  & Confidence & 7 (3) &  &  & Identifying a goal or purpose & 12 (7) \\
    &  & Determination & 2 (1) &  &  & Identifying known information & 13 (7) \\
    &  & Excited & 1 (1) &  &  & Identifying tasks & 29 (8) \\
    &  & Having fun & 1 (1) &  &  & Identifying unknown information & 4 (4) \\
    &  & Hopeful & 4 (2) &  &  & Improvement ideas & 33 (7) \\
    &  &  &  &  &  & Interpreting predictions & 8 (6) \\
    &  &  &  &  &  & Testing predictions & 12 (6) \\
    \hline
    \end{tabular}
\end{table*}    
     
There are 11 unique codes in the ``Carrying out the experiment'' theme in guided inquiry, with the most frequently referenced ones being ``Measurement comparison,'' ``Checking,'' and ``Visual observation.'' There are 14 unique codes in the ``Experimental process and components'' theme in open inquiry and the most frequent ones are ``Observations,'' ``Identifying measurements,'' and ``Measurement limitations.'' ``Observations'' include visual observations in the experiment as well as other general observations. ``Identifying measurements'' and ``Measurement limitations'' are unique in open inquiry, which indicates that this experiment successfully prompted the students to think often as well as deeply about the measurements they should make and the experimental limitations of their measurements. We did not see these student behaviours in the guided-inquiry experiment.

There are a total of 16 unique codes in the ``Interpersonal learning'' theme in guided inquiry, out of which 12 unique codes appeared in the ``Peer interaction'' category. The interactions in the guided-inquiry experiment were often between the two students performing the experiment and the students were ``Apologizing'' to each other, being in ``Agreement'' with one another, ``Doubting or checking,'' ``Correcting,'' ``Directing,'' ``Expressing thinking,'' making ``Observation,'' and ``Reconciling understanding.'' This is consistent with the students more frequently ``Informing'' and ``Suggesting'' to each other in the guided-inquiry experiment, shown in the common codes table~\ref{tab:commoncodes}. There are a total of 21 unique codes in the ``Self and interpersonal experiences'' theme in open inquiry, out of which 19 unique codes appeared in the ``Interpersonal interactions'' category. The most frequent ones include ``Instructor answering questions,'' ``Instructor checking in,'' ``Instructor guidance,'' ``Instructor prompting,'' and students seeking ``Clarification'' and ``Helping one another.'' We saw richer and more in-depth interactions in the open-inquiry experiment between the instructor and students, and between students in different groups.

There are 8 unique codes in the ``Sense making'' theme in guided inquiry including ``Confirmation'' and ``Procedure questions'' shown in table~\ref{tab:Unique2}. In open inquiry there are a total of 10 unique codes in the ``Comprehension experience'' theme, including ``Reflection on design or process'' and ``Reflection on laboratory experience'' in the ``Learning'' category. Here is an example of ``Reflection on design or process'':
\begin{quote}\vspace*{-0.05in}
A: \textit{Well the other one is still valid though. It's still a valid method, we're just taking the next one, which makes sense why it's not exactly.}\\
B: \textit{No no, I feel like, well our L is technically, our L is actually what we wrote as d. L, it seems like it's the distance.}
\end{quote}\vspace*{-0.05in}
And an example of ``Reflection on laboratory experience'' is:
\begin{quote}\vspace*{-0.05in}
A: \textit{It really doesn't seem like it's gonna be very long.}\\
B: \textit{No [be]cause I think a lot of like the lab is spent like kind of figuring out what to do.}\\
A: \textit{Yeah.}\\
B: \textit{So then your actual experiment is like if you came back in and knowing exactly what to do.}
\end{quote}\vspace*{-0.05in}

Finally, there are 12 unique codes in the ``The affective experience'' theme in guided inquiry all in the ``Emotions'' category except ``Relevance of physics.'' The negative emotions include ``Careless,'' ``Frustations,'' ``Making mistakes (disappointed),'' ``Sarcasm,'' ``Self doubt,'' and ``Stress.'' The positive emotions are ``Confidence,'' ``Determination,'' ``Excited,'' ``Having fun,'' and ``Hopeful.'' The most frequent emotion the students experienced was ``Frustration,'' for example,
\begin{quote}\vspace*{-0.05in}
\textit{We should have done the green and like the adjusting, then we should have done the red, why didn't they just tell us to do that. You know what I mean?}
\end{quote}\vspace*{-0.05in}
We did not see any of these emotions in open-inquiry experiment. Instead, we saw many unique codes in the ``Experimental design experience'' theme including ``Critical thinking'' and 15 codes in the ``Design'' category. Students did some ``Decision making'':
\begin{quote}\vspace*{-0.05in}
\textit{Okay so 10 degrees \dots What should we do, 10? Do something a bit more. Let's do a 5 degree one as well \dots}
\end{quote}\vspace*{-0.05in}
They were ``Identifying tasks'':
\begin{quote}\vspace*{-0.05in}
A: \textit{We need \dots}\\
B: \textit{Oh the angle measurement, that's the other thing we need.}\\
A: \textit{Angle between that and the wall?}\\
B: \textit{Well we can do the angle between this and the vertical.}
\end{quote}\vspace*{-0.05in}
The students were talking about ``Improvement ideas'': 
\begin{quote}\vspace*{-0.05in}
    \textit{Um maybe it, maybe the red [laser] will be better, or maybe if it was closer [than] we can see.}
\end{quote} \vspace*{-0.05in}
And ``Next steps'':
\begin{quote}\vspace*{-0.05in}
    \textit{We're gonna be able to get a bunch of data. And we're gonna know like for each data point like n, the wavelength, $\sin\theta_n$, and $\sin\theta_0$, and then that's what's gonna give us d.}
\end{quote}\vspace*{-0.05in}

\vspace*{-0.15in}
\subsection{Mapping onto Learning Outcomes and AAPT's Guidelines}
\vspace*{-0.15in}
Our findings are consistent with the intended laboratory learning outcomes we outlined in table~\ref{tab:outlineComp}. Specifically, both the guided- and open-inquiry experiments asked the students to align their apparatus, make measurements and calculations, and estimate uncertainties. These learning outcomes are reflected in the common codes that are present in both experiments: ``Making adjustments,'' ``Using equipment,'' ``Making measurements,'' ``Calculations,'' and ``Estimating uncertainty.'' In addition, the open-inquiry experiment prompted the students to make predictions, decide what physical quantities to measure, and design experimental procedures. These were reflected in the 27 references in the ``Making predictions'' code, 64 references in the unique ``Identifying measurements'' code, and a total of 222 references in the 15 unique codes in the ``Experimental design experience'' theme, respectively.

Furthermore, the AAPT guidelines recommend that students should be able to design a procedure to make a measurement, should have a hands-on opportunity to construct an apparatus, should do basic troubleshooting, should understand the limitations of their experimental design, and reflect on their results and suggest ways to improve their design. We saw in our analysis results that, while the guided-inquiry experiment addressed some of these recommended learning outcomes, the open-inquiry experiment allowed a much deeper and broader coverage of all aspects of these recommendations.

\vspace*{-0.15in}
\section{Conclusion}
\vspace*{-0.15in}
In this study we analyzed audio transcripts of students' conversations that took place while they were performing either a guided-inquiry experiment, in which procedures were provided, or an open-inquiry experiment, in which students were required to design their own procedures. We compared student learning experience and behaviour for both inquiry levels, by studying the results of qualitative and quantitative analysis of the transcripts, and found many differences.

First, we found that students in the guided-inquiry experiment, with the ultimate goal of completing all the prescribed tasks, focused on following the detailed instructions that were provided as evidenced by their frequent referencing of the lab manual. In contrast, the open-inquiry experiment provided the students with the freedom to explore a range of approaches and design their own experimental procedures. 
 
Second, we saw richer and more in-depth interactions in the open-inquiry experiment. Interactions during guided inquiry were generally between students who asked each other questions and told each other what to do. In open inquiry, there were more interactions between the instructor and students, and between students in different groups, and the conversations focused more on procedural design. 

Third, students in guided inquiry expressed many emotions, often negative ones including frustration and confusion. This was somewhat surprising since the students were given detailed procedures in the lab manual and should know very well what to do in the next steps. We argue that the negative emotions could be correlated with our first finding that the students came into the laboratory with the mindset that they would be able to carry out the experiment by simply following the instructions. When they experienced technical or other unexpected problems, they often reverted to express negative emotions including frustration and confusion. In comparison, students in open inquiry did not express any frustrations and they showed confusion less frequently. These students seemed to have come into the laboratory knowing that the lab manual would not give them all the answers and they were expected to figure out the next steps. With this mindset, these students in general had more positive experiences than those in guided inquiry.

Finally, the students in open inquiry had more opportunities to develop their experimental design skills. Many unique learning experience and behaviour emerged in the ``Comprehension experience'' and ``Experimental design experience'' themes, which indicate a richer and more comprehensive design and learning experiences for these students. Although the open-inquiry experiment did not ask the students to design all aspects of this experiment, the tasks  required the students to develop skills in one of the core AAPT curriculum areas, experimental design. It provided an opportunity for students to configure their apparatus, troubleshoot their apparatus and method, reflect on their results, and evaluate their procedure, and consequently they started to think like a physicist.

We find that student learning experience and behaviour in physics undergraduate laboratory experiments can be significantly improved by increasing the level of inquiry from guided to open. In some cases, the experimental apparatus does not have to be changed; it is sufficient to replace recipe-like procedures with questions or prompts that give students the freedom to design some of the experimental activities themselves. Alternatively, students can be given an experimental problem or task and the apparatus can be configured to accommodate a variety of experimental approaches. 
Using the analogy from mechanics introduced above, these experiments can be considered to have multiple DOF from which the students have to choose one. In the experiment described here, students had to design a strategy that resulted in the best experimental precision they could achieve, after a consideration of the experimental uncertainties and the available approaches.

The increased level of inquiry promotes AAPT learning outcomes and has the beneficial side-effect of reducing negative affective experiences for students. An important finding being that detailed experimental procedures can, counter-intuitively, be the source of the negative affective experiences.

We suggest that when instructors are designing open-inquiry laboratory experiments or transforming a guided-inquiry experiment into an open-inquiry experiment, they consider whether instructions could be turned into questions or prompts. Rather than telling the students what to do, students can be given the freedom to evaluate a number of experimental options. Clearly, for this to be successful, the apparatus has to support more than one experimental strategy. It is not always necessary to ask the students to design a complete experimental procedure to raise the inquiry level. Nevertheless, the instructor has to find a manageable balance between the amount of detail given in the lab manual and the number of design tasks that the students are required to execute.

This study describes the positive effect of increasing the inquiry level of undergraduate physics experiments on student behaviour and learning. Moreover, it provides guidelines on how to best design or redesign undergraduate experiments to support open-inquiry, which we have posited as a way of transferring agency from the instructor to the student. However, we do recognize that a study of this sort has limitations that, for completeness, we delineate here: 
1. It was based on a specific experimental topic on CD diffraction; 2. The students in this study were from a second-year undergraduate laboratory course in one research institution, therefore may not represent those who are new to universities or more senior in their undergraduate studies; 3. The students who did the guided-inquiry and open-inquiry experiments are different and randomly selected from the class, which may not represent a general population; and 4. While we categorized our experiments on two specific inquiry levels, we recognize that each experiment remains unique. The design of each experiment must retain some degree of flexibility to support student familiarity with the topic and expertise with a specific experimental process. As a result, the level of instructor guidance, in terms of the number and type of written lab manual and in-lab prompts, may shift accordingly. 

\begin{acknowledgments}
This work was funded by: the TRESTLE network~\cite{trestlewebsite} (NSF DUE1525775); the Department of Physics, Engineering Physics and Astronomy; the Center for Teaching and Learning; and the Faculty of Engineering and Applied Science at Queen's University. 
We would like to thank: C. Knapper, R. Knobel, G. L. McLean and M. Swarthout for critical readings of this manuscript. 
\end{acknowledgments}

\bibliography{Cai_et_al_Designing_Experiments}
\end{document}